\documentclass[aps,amsmath,amssymb,preprintnumbers,groupedaddress,twocolumn]{revtex4}
\usepackage{graphicx,psfrag,hyperref}

\def\hepth#1{\href{http://xxx.arxiv.org/abs/hep-th/#1}{{arXiv:hep-th/#1}}}

\def\math#1{\href{http://xxx.arxiv.org/abs/math/#1}{{arXiv:math/#1}}}

\def\arxiv#1#2{\href{http://xxx.arxiv.org/abs/#1}{{arXiv:#1 [#2]}}}

\newcommand{\beq}{\begin{equation}}
\newcommand{\eq}{\end{equation}}
\newcommand{\req}[1]{(\ref{#1})}
\newcommand{\ie}{{\it i.e.}}

\newcommand{\Fcal}{\mathcal F}
\newcommand{\Gcal}{\mathcal G}

\newcommand{\Ncal}{\mathcal N}
\newcommand{\Mcal}{\mathcal M}

\newcommand{\calm}{\mathcal M}

\newcommand{\calg}{\mathcal G}

\renewcommand{\P}{\mathbb P}

\newcommand{\Z}{\mathbb Z}
\newcommand{\R}{\mathbb R}

\newcommand{\C}{\mathbb C}

\newcommand\topa[2]{\genfrac{}{}{0pt}{2}{\scriptstyle #1}{\scriptstyle #2}} 

\begin{document}

\preprint{CERN-PH-TH/2010-231, UCB-PTH-10/19}
\title{Shift versus Extension in Refined Partition Functions}

\author{Daniel Krefl$^{a}$ and Johannes Walcher$^{b}$}
\affiliation{\it ${}^{a}$ Center for Theoretical Physics, University of California, Berkeley, USA \\
\it $^{b}$ PH-TH Division, CERN, Geneva, Switzerland
}

\begin{abstract}
We have recently shown that the global behavior of the partition function of $\Ncal=2$ gauge
theory in the general $\Omega$-background is captured by special geometry in the guise of the
(extended) holomorphic anomaly equation. We here analyze the fate of our results under the shift of the mass parameters of the gauge theory. The preferred value of
the shift, noted previously in other contexts, restores the $\Z_2$ symmetry of the instanton
partition function under inversion of the $\Omega$-background, and removes the extension. 
We comment on various connections.
\end{abstract}
\maketitle
\noindent

\section{Introduction}

In recent work \cite{Krefl:2010fm}, we have initiated the systematic study of refined
(topologial string, or gauge theory in the general $\Omega$-background) partition functions
from the point of view of the special geometry governing the underlying moduli space.

Starting from the explicit expressions for the gauge theory partition function $Z(a,m,\epsilon_1,
\epsilon_2;q)$ which was obtained in \cite{nekrasov} using localization on the moduli space of instantons,
and parametrizing the $\Omega$-background according to
\beq\label{para}
\epsilon_1=\beta^{1/2}\lambda,\,\, \epsilon_2=-\beta^{-1/2}\lambda\,,
\eq
we expanded (at small $q$)
\beq
\label{valid}
\log Z(a,m,\epsilon_1,\epsilon_2;q) = \sum_{n=-2}^\infty \lambda^n \calg^{(n)}
(a,m,\beta;q)
\eq
Here, $a$ are the vectormultiplet moduli, $m$ parameterizes the masses of flavor
hypermultiplets, $\epsilon_1,\epsilon_2$ are the equivariant parameters of the 
$\Omega$-deformation and $q$ is the instanton counting parameter.

The expansion (\ref{valid}) is, initially, valid in the weak coupling region ($a\gg q$) of the 
Seiberg-Witten moduli space $M$. Following \cite{hukl1}, one can promote the $\calg^{(n)}(a,
m,\beta)$ to global, albeit non-holomorphic, objects defined over all of $M$. The main result
of \cite{Krefl:2010fm}, which generalizes the results of \cite{hukl1}, is that the non-holomorphic 
dependence of the $\calg^{(n)}$ is controlled by the holomorphic anomaly equations familiar from the 
topological string, for the general value of $\beta$ (the usual relation to topological string being 
recovered at $\beta=1$, \ie, $\epsilon_1+\epsilon_2=0$). Moreover, the holomorphic ambiguity can be 
completely fixed, order by order, by the ($\beta$-dependent) singularity structure of the $\calg^{(n)}$, 
using local canonical coordinates at each boundary of $M$.

A remarkable feature of the results in \cite{Krefl:2010fm} was the necessity to resort to the 
extended holomorphic anomaly of \cite{extended,tadpole}. This requirement was apparent
in the expansion (\ref{valid}), which generally goes also over odd powers of $\lambda$, and
hence does not fit into the standard framework in which the $\calg^{(n)}$ are identified with 
topological string amplitudes $\Fcal^{(g)}$ and $n=2g-2$ is even. On the other hand, the existence of
the odd sector in the expansion (\ref{valid}) appears to be in conflict with the refined BPS 
expansion of the topological string partition function proposed in \cite{Iqbal:2007ii}. 
Indeed, a quick peek reveals that that expansion is manifestly symmetric under $\lambda\to-\lambda$, 
{\it i.e.}, expressed in terms of the $\Omega$-background
\footnote{After completion of \cite{Krefl:2010fm}, we confirmed that the refined BPS invariants 
of local Calabi-Yau background such as local $\P^2$ (which is not directly a geometric engineering 
situation, but related to one via a blow-down) found in \cite{Iqbal:2007ii} can also be computed in 
our B-model scheme, using the holomorphic anomaly with vanishing extension. 
%Especially, the singularity structure at the conifold point is also governed by the $
%c=1$ string at $R=\beta$.  
This result was
also shown in M.~x.~Huang and A.~Klemm,
  %``Direct integration for general Omega backgrounds,''
  \arxiv{1009.112}{hep-th}.
  %%CITATION = ARXIV:1009.1126;%%
.},
we have the symmetry
\beq
\label{symmetry}
(\epsilon_1,\epsilon_2)\to (-\epsilon_1,-\epsilon_2)
\eq
and hence we should have found $\calg^{(n)}=0$ whenever $n$ is odd. The purpose of this note is to
release some tension about this point.

In \cite{Iqbal:2007ii}, the symmetry (\ref{symmetry}) was ensured by exploiting the redefinition
of flat coordinates (K\"ahler parameters $t$ of the A-model) that vanishes in the unrefined limit, 
schematically $t\to t+\delta t$ with $\delta t\propto (\epsilon_1+\epsilon_2)$. 
As we shall see below, we can in fact restore the symmetry, and remove the extension of the
holomorphic anomaly, also in the gauge theory case by shifting the mass parameters 
\beq
\label{crucial}
m\to m+(\epsilon_1+\epsilon_2)/2\,.
\eq
This is similar to the shifts in \cite{Iqbal:2007ii}, and is in fact related to them via 
geometric engineering, but differs from them in one crucial respect. Namely, while the 
variables being shifted in \cite{Iqbal:2007ii} are dynamical fields (moduli), the $m$ appear 
as external parameters in the gauge theory. Anticipating some of our conclusions, this means
that the information about this {\it shift of an a priori non-dynamical parameter gets traded 
in our formalism with the non-trivial extension} of the holomorphic anomaly equation. We will 
elaborate on this insight below.

The redefinition of mass parameters (\ref{crucial}) has played a significant r\^ole in the 
recent (``AGT'') relations between four and two-dimensional conformal field theories, see in 
particular \cite{agt,pestun}. That the instanton partition function should be invariant 
under the symmetry (\ref{symmetry}) was especially emphasized in \cite{nakajima}. We learned
in \cite{pestun} that shifts as in \req{crucial} first appeared in \cite{shadchin}.

\section{Instanton counting} 

According to \cite{ss}, \cite{nekrasov}, the instanton partition function in $\Omega$-background for 
gauge theory with $N_f$ flavors is given by 
\beq
\label{bundle}
Z^{\rm inst}(a,m,\epsilon_1,\epsilon_2;q) = \sum_k q^k \int_{\Mcal_k} {\bf e}(V\otimes S)
\eq
Here, $\calm_k$ is the (compactified) moduli space of instantons of charge $k$ on $\R^4$, $V$ is the
bundle of solutions of the Dirac equation over it, and $S\cong \C^{N_f}$ is the  flavor space.
The integration takes place in the equivariant cohomology $\C(a,m,\epsilon_1,\epsilon_2)$.
It was further shown in \cite{nekrasov} that at a fixed point labelled by a collection of partitions 
$Y=(Y_1,Y_2)$ (We will here consider only $SU(2)$ gauge theory with $N_f<4$ fundamental flavors.), 
the Euler class ${\bf e}(V\otimes S)$ localizes to
\begin{multline}
\label{wrong}
f^Y(a,m,\epsilon_1,\epsilon_2)=  \\ 
+ \prod_{\topa{k=1,..,N_f}{\gamma=1,2}} \prod_{s\in Y_\gamma} 
\bigl(a_\gamma + m_k + \epsilon_1(i-1)+\epsilon_2(j-1) \bigr)\,,
\end{multline}
where $i,j$ are the coordinates of the box $s\in Y_\gamma$, and $a_1+a_2=0$ is implicit.
The formula \req{wrong} was also found in \cite{complain2}. 

By combining the remarks in \cite{nakajima} and the original observations in \cite{shadchin} 
with the results of \cite{pestun}, those of \cite{agt}, and the methods of \cite{Iqbal:2007ii}, 
we are led to consider the alternative expression 
\begin{multline}
\label{alternate}
\tilde f^Y(a,m,\epsilon_1,\epsilon_2) = \\
= \prod_{\topa{k=1,..,N_f}{\gamma=1,2}} \prod_{s\in Y_\gamma} 
\bigl(a_\gamma + m_k + \epsilon_1(i-1)+\epsilon_2(j-1) +\textstyle{\frac{\epsilon_1+\epsilon_2}2} \bigr)\,.
\end{multline}
Several explanations and remarks are in order. First of all, we observe that we may obtain 
$\tilde f$ from $f$ by shifting the masses diagonally, 
\beq
\label{shift}
m\to  m+(\epsilon_1+\epsilon_2)/2\,.
\eq 
Now, as mentioned above, the redefinition \req{shift} is related to the procedure of 
\cite{Iqbal:2007ii} for ensuring a sensible BPS expansion of the refined topological string 
partition function. Second, in \cite{agt}, shifts by $(\epsilon_1+\epsilon_2)/2$ as in 
\req{shift}
were used to relate the mass parameters and Coulomb moduli appearing in the instanton 
partition function with the Liouville momenta labelling a conformal block dual to the 
instanton partition function \footnote{It is interesting to remark here that AGT found it
expedient to divide the gauge theory partition function by a certain $U(1)$ factor,
in addition to restricting to $a_1+a_2=0$. This decoupling plays no apparent r\^ole 
in our discussion.
}. Next, the relation \req{shift} between the mass parameters of \cite{agt} 
(which are those of \cite{nekrasov}) and those of the localization computation of 
\cite{pestun0} for the $\Ncal=2^*$ gauge theory on $S^4$ was found in \cite{pestun} to 
ensure the restoration of $\Ncal=4$ supersymmetry in the limit $m\to 0$. (Note that this is the 
mass of an adjoint hypermultiplet.) This was interpreted in \cite{pestun} as the ``physical'' 
definition of the mass parameter.

Finally, in \cite{nakajima}, a computation showed that it would be more natural to use the kernel 
of the Dirac operator instead of the Dolbeault operator coupled to the instanton background in 
\req{bundle} as the definition of the gauge theory with fundamental matters, \ie, to twist the 
fermions. As usual, this is accomplished by tensoring with the half canonical bundle, a line 
bundle with weight $(\epsilon_1+\epsilon_2)/2$. In the localization, this is precisely equivalent 
to the shift \req{shift} of the mass parameters.

For comparison, we find it convenient to interpolate between the two prescriptions \req{wrong} 
and \req{alternate} by introducing an additional parameter $\xi$ continuously tuning the 
magnitude of the shift. Thus, we consider the family of partition functions
\beq
\label{zetashift}
Z_\xi(a,m,\epsilon_1,\epsilon_2;q)=Z_1(a,m+(1-\xi)\textstyle{\frac{\epsilon_1+\epsilon_2}{2}},
\epsilon_1,\epsilon_2;q))\,.
\eq
Here $\xi=1$ corresponds to using \req{wrong}, as in \cite{Krefl:2010fm}. The value $\xi=0$
corresponds to \req{alternate}. In general, we find that $Z_0$ is symmetric (for all values of
$m$) under $(\epsilon_1,\epsilon_2)\to (-\epsilon_1,-\epsilon_2)$, as announced in 
\cite{nakajima}. We also find that the theory is invariant under $\xi\to -\xi$.

\section{B-model}

It was found in \cite{Krefl:2010fm} that the amplitudes $\calg^{(n)}$ defined as coefficients 
of $\lambda^n$ in the expansion \req{valid} of $Z_1$ (for $SU(2)$, $N_f=0,1,2,3$ flavors, with
$m\equiv 0$) satisfy, when appropriately continued to modular invariant expressions over 
the Seiberg-Witten moduli space, the extended holomorphic anomaly equation of 
\cite{extended,tadpole}. The Griffiths infinitesimal invariant measuring the extension vanishes 
for $N_f=0,2,3$, and for $N_f=1$ can be obtained from the chain
integral of the Seiberg-Witten differential between an appropriate pair of points on the
Seiberg-Witten curve. The singularity structure around monopole/dyon points was enough to
completely fix the holomorphic ambiguity.

To give a bit more details, we recall that the full gauge theory partition function $Z_\xi$
is the product of the instanton part $Z_\xi^{\rm inst}$, discussed above, and a perturbative 
part $Z^{\rm pert}_\xi$. To write this piece, we introduce as in \cite{Krefl:2010fm} the two 
sets of functions $\Phi^{(n)}(\beta)$ and $\Psi^{(n)}(\beta)$ by the asymptotic expansion of 
the two Schwinger integrals
\beq
\int \frac{ds}{s}\frac{e^{-x s}}{(e^{\epsilon_1 s}-1)
(e^{\epsilon_2 s}-1)}\sim \dots+\sum_{n>0}\frac{\lambda^n}{x^n}\Phi^{(n)}(\beta)\
\eq
%and
\beq
\int \frac{ds}{s}\frac{e^{-x s} e^{(\epsilon_1+\epsilon_2)s/2} }{(e^{\epsilon_1 s}-1)
(e^{\epsilon_2 s}-1)}\sim \dots+\sum_{n>0}\frac{\lambda^n}{x^n}\Psi^{(n)}(\beta)\,.
\eq
(The $\Phi^{(n)}$ are essentially the $\gamma_{\epsilon_1,\epsilon_2}$ of \cite{nekrasov,neok,nayo2},
and the $\Psi^{(n)}$ appear as $\delta_{\epsilon_1,\epsilon_2}$ in \cite{nakajima}.)
We then have for vanishing bare mass of the fundamentals \cite{agt,nakajima}
\beq
\label{clear}
\log Z^{\rm pert}_0 \sim \sum_{n\; {\rm even}} 
{\lambda^n}\Bigl(\frac{2\Phi^{(n)}(\beta)}{(2a)^n} -\frac{ 2 N_f \Psi^{(n)}(\beta)}{a^n}\Bigr)\,.
\eq
The first term comes from integrating out the $2$ vectormultiplets (W-bosons) of BPS mass $\pm 2a$ 
in the limit $a\to\infty$, and the second term from the $2N_f$ hypermultiplets of mass $\pm a$. 

Similarly, the leading behaviour around a point with massless monopole/dyon is governed by 
$\Psi^{(n)}(\beta)$, corresponding to integrating out a light hypermultiplet with mass 
given by the local flat coordinate. 

We may now repeat the calculations of \cite{Krefl:2010fm} (which were done for $\xi=1$), for 
general value of $\xi$. We find that the amplitudes $\Gcal^{(n)}_\xi(\beta)$ appearing in the 
expansion of $\log Z_\xi$ are always governed by the extended holomorphic anomaly equation. In 
particular, we find that imposing the $\beta$- and $\xi$-dependent gap structure at the monopole/dyon 
points completely fixes the leading weak coupling behaviour given by \req{clear} (and its shifts).
As observed above, the $\Gcal^{(n)}$ of $n$ odd vanish for $\xi=0$, so in this 
case we use the standard holomorphic anomaly equation, even for $N_f=1$, and $3$.

Let us also briefly comment on the case with two flavors. As noted in \cite{Krefl:2010fm} at $\xi=1$,
the $\Z_2$ symmetry between monopole and dyon point (which obtains in the unrefined case 
$\beta=1$), is broken for generic value of $\xi$. The explicit calculation shows that this
symmetry is in fact restored at $\xi=0$, giving additional corroboration that this is the
most symmetric value. Correspondingly, at this value of $\xi$, the leading singularities at
monopole and dyon points are both captured by the $\Psi^{(n)}$ coefficients.

Finally, we emphasize again that we have performed these calculations only for vanishing
bare mass of the flavor hypermultiplets, $m=0$. It would be interesting to check the massive 
case as well, and in particular take a look at the various superconformal points in the space 
of theories.

\section{Discussion}

In this brief note, we have scouted the freedom of shifting the masses of fundamental 
hypermultiplets of $\Ncal=2$ supersymmetric $SU(2)$ gauge theory by the self-dual 
$\Omega$-background parameter, $\epsilon_1+\epsilon_2$. We have seen that for all values of the 
shift parameter $\xi$, the deformed partition function $Z_\xi$ is controlled in the B-model
by the extended holomorphic anomaly equation together with appropriate boundary conditions.
The value $\xi=0$ is prefered by the circumstance that $Z_0$ is symmetric in $\lambda\sim
\sqrt{\epsilon_1\epsilon_2}$, and that the B-model formalism reproduces the perturbative 
spectrum most precisely. This is also the value of the shift for which the extension vanishes.
One may wish to conclude at this point. However, the consistency of the results for $\xi\neq 0$ 
(especially, $\xi=1$) \footnote{There is at least one way in which $\xi=1$ is also special. 
Namely, the orientifold of the $N_f=2$ theory with $\xi=1$, $\beta=1$ and appropriately chosen 
masses naturally yields the same theory without orientifold but at $\beta=2$ \cite{Krefl:2010fm}.}, 
the naturalness of the extension, and general curiosity begs the question: Is there a physical 
meaning of the shift?

We can obtain some first hints about this question by taking a higher-dimensional perspective,
\ie, by embedding the gauge theory in string theory using geometric engineering \cite{geoeng}.
For simplicity, let us consider the $N_f=1$ theory. Then the relevant geometry is a Hirzebruch
surface with attached conifold-like geometry (see for instance \cite{hollywood}). In particular, 
it has three
K\"ahler parameter $Q_i=e^{-t_i}$, $i=1,2,3$. The first of these (the size of the base) 
controls the geometric engineering decoupling limit, the other (the size of the fiber) becomes 
identified with the Coulomb modulus $a$ in this limit, and the last one (the size of the attached 
conifold), is the mass parameter $m$ of the gauge theory. Note that the field theory vev and the 
mass parameter have a common geometric origin as closed string moduli.

According to the conjectures originating in \cite{nekrasov}, the gauge theory partition
function in general $\Omega$-background should correspond in the string theory to an 
appropriate ``refinement'' \cite{hollywood} of the topological string amplitudes. 
According to \cite{hollywood,Iqbal:2007ii}, the spacetime interpretation of this refined 
topological string should capture BPS state counting taking account of the spin. As of this 
writing, there is no compelling proposal for the definition of this refined topological string, 
neither from worldsheet nor from target space field theory in either A- or B-model. 
In the A-model, however, 
we have the ``refined topological vertex'' \cite{Iqbal:2007ii}
that allows the computation of refined amplitudes precisely for geometries
that engineer $\Ncal=2$ gauge theories. In fact, this refined vertex was constructed 
precisely to match the five-dimensional version of the instanton partition function obtained
in \cite{nekrasov,neok}. An important aspect of the formalism is the exploitation
of the freedom to shift the K\"ahler parameters before identifying them with 
physical quantities such as the masses of (refined) BPS states, or field theory
vectormultiplet moduli. The fixing of these shifts could provide important hints
for completing the refined vertex formalism.

In the engineering setup, of course, the shift of the mass parameter of the gauge 
theory $\delta m\propto \epsilon_1+\epsilon_2$ lifts in the string theory to the shift of the
K\"ahler modulus $\delta t_3\propto\epsilon_1+\epsilon_2$. This is precisely the type of
shift utilized in \cite{Iqbal:2007ii} to ensure that there is a BPS state counting interpretation. 
It is not hard to check that our preferred value of the shift at $\xi=0$ lifts precisely to an even 
sector only (integer) refined BPS state counting for the geometry engineering $N_f=1$ theory.
However, we stress that $Z_\xi$ generally fulfills the extended holomorphic anomaly 
equation. Only for the specific value $\xi=0$ does it reduce to the standard holomorphic 
anomaly equation. Also, we note that one can obtain the full family $Z_\xi$ as the effective field 
theory limit of the refined topological vertex partition functions on the corresponding geometric 
engineering geometry, with the shift of mass lifting to a shift of the corresponding K\"ahler 
modulus.

In the topological string, the shift of $t_3$ for non-zero values of $\xi$ leads to a non-trivial
odd sector with $\Gcal^{(-1)}\sim \xi \partial_{t_3}\Fcal^{(0)}$, with $\Fcal^{(0)}=\Gcal^{(-2)}$ 
the standard prepotential. This is of course nothing but a closed string period, and hence leads 
to a vanishing extension in the holomorphic anomaly equation. This leads to the following speculative 
interpretation of the shift \footnote{We thank M.\ Aganagic for pointing us towards this direction}:
In the (to be refined) topological A-model we are shifting the K\"ahler parameters as
\beq
\label{closer}
t\rightarrow t+N\lambda \,,
\eq
with $N$ some number, and $\lambda$ the topological string coupling constant. In the (to be refined)
mirror B-model this looks as though switching on $N$ units of flux through the corresponding 3-cycle,
cmp.\ \cite{mina}. 
Hence, it might be possible to interpret $Z_\xi$ as the effective field theory limit of the partition 
function of the engineering geometry with additional flux switched on, or, in the spirit of flux/brane 
duality, with additional D-branes.

Taking the decoupling limit, the closed string K\"ahler parameter $t_3$ becomes the non-dynamical 
mass parameter $m$. While it is still true that the shift introduces an odd sector with $\Gcal^{(-1)}
\propto \xi \partial_m\Fcal^{(0)}$, the latter is no longer a closed period, and hence gives rise, 
in general, to a non-trivial extension \footnote{One may derive the pair of points obtained in 
\cite{Krefl:2010fm} by studying the deformation of the Seiberg-Witten curve under turning on the 
masses.}. As a result, we can no longer interpret the shift in terms of a closed string flux.
However, we may still interpret it in terms of adding $N$ background D-branes, very much
in the spirit of the original purpose of the extended holomorphic anomaly equation \cite{extended}.
In fact, keeping track of the ``number of boundaries'' via the independent ``open string'' coupling 
constant $\propto\xi\lambda$ should allow the reconstruction of the full $m$-dependence. 

This type of reasoning might become more compelling if we leave aside the gauge theory 
interpretation, and only focus attention on the structure of the holomorphic anomaly
equation and its solutions, following the line of investigation initiated in 
\cite{ooguri,newa}. Thus, we view the solution of the extended holomorphic anomaly 
abstractly as an open-closed string wavefunction, with the extension specifying the
D-brane background, and two independent parameters $(\lambda,\xi)$ playing the r\^ole
of closed and open string coupling, respectively. In this language, the main lesson of our discussion is that the open-closed wavefunction
$Z(\lambda,\xi;a,m)$ is equal to a purely closed string wavefunction $\tilde 
Z(\lambda;a,m+\delta m)$ with a shift $\delta m \propto \xi$ of the ``non-dynamical closed 
string'' field $m$.  

We emphasize that this type of open-closed string relation is different from the one
first proposed in \cite{ooguri}, also in the context of the extemded holomorphic anomaly 
equation. As explained in \cite{newa}, the improved shift of \cite{ooguri} does remove the
extension at the level of the holomorphic anomaly equation, but does not reconstruct
the known purely closed string (at least not in a recognizable form). See also \cite{italians} 
for further discussion of the shift of \cite{ooguri} in the light of open-closed string 
correspondence \footnote{The point of view on open-closed relation taken here is closer to the 
flux shift \req{closer}, and was also proposed a while ago by C.\ Vafa, and S.\ Shatashvili.}.

\enlargethispage{5cm}
 
\acknowledgments
We thank Mina Aganagic, Amer Iqbal, Can Koz\c caz, Wolfgang Lerche, Peter Mayr, Sara Pasquetti, 
and Samson Shatashvili for valuable dicussions and 
comments. J.W.\ thanks the Departments of Physics and Mathematics at McGill University, and 
the Simons Workshop on Mathematics and Physics, 2010, for hospitality during the course of this 
work. The work of D.K. was supported in part by a Simons fellowship and by the WPI initiative by 
MEXT of Japan.

%%%%%%%%%%%%%%%%%%%%%%%%%%%%%%%%%%%%%%%%%%%%%%%%%%%%%%%%%%%%%%%%%
%%%
%%%                     BIBLIOGRAPHY
%%%
%%%%%%%%%%%%%%%%%%%%%%%%%%%%%%%%%%%%%%%%%%%%%%%%%%%%%%%%%%%%%%%%%

\end{document}